\documentclass[11pt]{article}
\usepackage{times}
\usepackage{graphicx}
\usepackage{url}
\usepackage{mathfont}
\def\log{\mathop{{\rm log}}}

\mathcode`O="724F

\makeatletter
\def\@begintheorem#1#2{\it \trivlist \item[\hskip \labelsep{\bf #1\ #2.\
}]}
\def\@opargbegintheorem#1#2#3{\it \trivlist
      \item[\hskip \labelsep{\bf #1\ #2\ (#3).}]}
\makeatother

\newtheorem{theorem}{Theorem}
\newtheorem{defn}{Definition}
\newtheorem{lemma}{Lemma}

\newcommand{\qed}{~$\vrule width.15cm height.2cm depth0cm$ \medbreak}
\newenvironment{proof}{\noindent{\bf Proof sketch: }}{\qed}

\DeclareSymbolFont{AMSb}{U}{msb}{m}{n}
\DeclareSymbolFontAlphabet{\Bbb}{AMSb}
\def\R{\ensuremath{\Bbb R}}

\def\Flat{{\cal F}}
\def\Halfspace{{\cal H}}
\let\Hyperplane\Halfspace
\let\Hyper\Halfspace
\def\Sphere{{\cal S}}
\def\Line{{\cal L}}

\newcommand{\OCfigure}[3]{\begin{figure}
 \center{\vbox to #3 {\vfil
	 \includegraphics[height=#3]{#1.eps}}}
 \caption{#2}
 \label{#1}
\end{figure}}

\begin{document}

\title{Computing the Depth of a Flat}
\author{Marshall Bern\thanks{Xerox PARC, 3333 Coyote Hill Rd., Palo 
Alto, CA 94304}\and
David Eppstein\thanks{UC Irvine, Dept. Inf. \& Comp. Sci., Irvine, CA 
92697. Work done in part while visiting Xerox PARC.}}
\date{ }

\maketitle

\begin{abstract}
We compute the regression depth of a $k$-flat in a set of $n$ points
$\R^d$, in time
$O(n^{d-2}+n\log n)$ when $1\le  k\le d-2$.  
In constrast,
the best time bound known for the $k=0$ case
(data depth) or for the $k=d-1$ case (hyperplane regression) 
is $O(n^{d-1}+n\log n)$.
\end{abstract}

\section{Introduction}

Regression depth was introduced by Hubert and
Rousseeuw~\cite{RouHub-JASA-99} as a distance-free quality measure for
linear regression. 
The depth of a hyperplane with respect to a set of data points
in $\R^d$ is the minimum number of
data points crossed in any continuous motion taking the hyperplane to a 
vertical  hyperplane.  A vertical hyperplane is 
a regression failure, because it allows the response variable 
(that is, the dependent variable)
to vary over its entire range while keeping the explanatory
variables (the independent variables) fixed.
 Thus a good regression plane should
be far from a vertical hyperplane. 
A deepest hyperplane is farthest from vertical in a combinatorial sense;
it provides a good fit even in the presence of skewed
or data-dependent errors, and is
robust against a constant fraction of arbitrary outliers.

Due to its combinatorial nature, the notion of regression depth leads 
to many interesting algorithmic and geometric problems.
For points on the line $R^1$, a median point is a point of maximum depth.
For the case of $n$ points in the plane $R^2$, 
Hubert and Rousseeuw~\cite{HubRou-JMA-98}
gave a simple construction called the {\em catline},
which finds a line of depth $\lceil n/3\rceil$. 
The deepest line in the plane can be found in time
$O(n\log n)$~\cite{LanSte-SODA-00}.
The catline's depth bound is best possible, 
and more generally in $\R^d$ the best depth bound is
$\lceil n/(d+1)\rceil$~\cite{AmeBerEpp-DCG-00,Miz-IMSB-98}. 
The fastest known
exact  algorithm for maximizing depth takes time $O(n^d)$, and 
$\epsilon$-cutting techniques can be used to obtain an $O(n)$-time
$(1+\epsilon)$-approximation to the maximum
depth~\cite{SteWen-CCCG-98}.

In previous work~\cite{BerEpp-SCG-00}, we generalized depth to
{\em  multivariate regression}, that is, fitting points in $\R^d$ by
affine  subspaces with dimension $k<d-1$ ({\em $k$-flats}
for  short).
We showed that for any $d$ and $k$, deep $k$-flats always exist,
meaning that for any point set in $R^d$,
there is always a $k$-flat of depth a constant fraction 
of $n$, with the constant depending on $d$ and $k$. 
This result implies that the deepest flat is robust,
with a breakdown point which is a constant fraction of $n$.
We also generalized the catline construction to find lines with
depth $\lceil n/(2d-1)\rceil$, which is tight for $d\le 3$  and would be
tight for all $d$ under a conjectured 
$\lceil n/((k+1)(d-k)+1)\rceil$ bound on maximum regression depth.
On the algorithmic side, we showed that
$\epsilon$-cuttings can be used to  obtain an $O(n)$-time
$(1+\epsilon)$-approximation for the deepest flat.

In this paper, we consider the problem of testing the depth of a given 
flat, or more generally the crossing distance between two flats.
Rousseeuw and Struyf~\cite{RouStr-SC-98} studied similar problems for
hyperplanes and points. 
The crossing distance between a point and a
hyperplane  can be found in time $O(n^{d-1}+n\log n)$ by examining the
arrangement's restriction to the hyperplane (as described later),
and the same bound applies to testing the depth of a 
hyperplane or point. We show that, in contrast, the depth of a flat 
of any other dimension can be found in randomized time
$O(n^{d-2}+n\log n)$.  More generally, the crossing distance between 
a $j$-flat and a $k$-flat can be found in time $O(n^{\,j+k-1}+n\log n)$
when $1\le j,k\le d-2$.

\section{Definitions}

A generic $k$-flat with $k < d-1$ can move continuously
to vertical without crossing any data points, so it is
not obvious how to generalize regression depth to $k$-flats.
The key is to start from an equivalent definition
of hyperplane regression depth:  the depth of a hyperplane $\Hyper$
is the minimum number of data points in a {\it double wedge\/}
with one boundary equal to $\Hyper$ 
and the other boundary vertical (parallel to the response variable's axis). 
A double wedge is the closed region bounded by two hyperplanes;
it is the region necessarily swept out 
by a continuous motion of one bounding hyperplane to the other.

Now consider the simplest example with $k < d-1$, the 
regression depth of a line in $\R^3$.
We think of $x$ as the explanatory variable,
and $y$ and $z$ as two response variables.
A regression line simultaneously explains $y$ and $z$ as linear
functions of $x$, and any line parallel to the $yz$-plane is 
a regression failure that allows $y$ and $z$ to vary over their
entire range while keeping $x$ fixed. 
We would thus like our regression line to be far from
lines parallel to the $yz$-plane.
A reasonable guess at the definition of regression
depth of a line $\Line$ would be the minimum number of data points 
in a double wedge with one boundary containing $\Line$
and the other boundary parallel to the $yz$-plane.

This guess indeed turns out to be the correct generalization; 
its naturalness is revealed by looking
at the dual formulation of the problem.
The projective dual of a point set is a hyperplane arrangement, and 
hyperplane regression dualizes to finding a central point in an 
arrangement.
If the {\em depth\/} of a point $p$ is the 
minimum number of arrangement hyperplanes crossed by any line segment 
from $p$ to the hyperplane at infinity, then (as observed
by Rousseeuw) the regression 
depth of a hyperplane is exactly the depth of its dual 
point in the dual arrangement.  
We generalized this observation to give a natural distance measure
between flats in an arrangement~\cite{BerEpp-SCG-00}.

\begin{defn}\label{crossdefn}
The {\bf crossing distance} between two flats in 
an arrangement is the fewest hyperplane crossings along any line segment
having one endpoint on each flat. 
\end{defn}

In the primal formulation,
the crossing distance is the minimum number of points
in a double wedge with one boundary containing one flat
and the other boundary containing the other.

\begin{defn}\label{depthdefn}
The {\bf regression depth of a $k$-flat} is the crossing distance
between its dual $(d-k-1)$-flat and a $k$-flat at {\em vertical infinity}. 
\end{defn}

For multivariate regression,
the $k$-flat at vertical infinity should be the one dual to
the intersection of the hyperplane at infinity
with the $(k-d)$-flat spanned by the response variable axes.
With this choice, regression failures have regression depth zero.
For hyperplane regression, there is no choice to make as there 
is only one $(d-1)$-flat at infinity.

Along with hyperplane regression,
Definition~\ref{depthdefn} also subsumes the classical notion
of {\em data depth} or {\em Tukey depth}.
The data depth of a point $p$ is the minimum
number of data points in any closed half-space---a degenerate
double wedge---containing $p$.
The data depth of $p$ is also the crossing distance of its
dual hyperplane from the point at vertical infinity.

\section{Reduction to Covering}

We now show that crossing distance can be reduced to finding a minimally
covered point in a certain family of sets.  Suppose we are
given an arrangement of hyperplanes, a $j$-flat $\Flat_{1}$, and a
$k$-flat $\Flat_{2}$. We wish to determine the line segment,  having one
endpoint on each flat, that crosses as few arrangement  hyperplanes as
possible.

We first parametrize the space of relevant line segments. 
Without loss of generality the two flats do not meet (else the crossing
distance is zero) so any  pair of points from $\Flat_1\times\Flat_2$
determines a unique line.   The pair divides the line into two 
complementary line segments (one through infinity), 
so we need to augment each point of  $\Flat_1\times\Flat_2$ by
an  additional bit of information to specify each possible line segment.
We do this topologically: $\Flat_1$ is a projective space, having as 
its double cover a $j$-sphere $\Sphere_1$, and similarly the double 
cover of  $\Flat_{2}$ is a $k$-sphere $\Sphere_2$.  The product 
$\Sphere_1\times\Sphere_2$ supplies two extra bits of information per 
point, and there is a continuous two-to-one map from 
$\Sphere_1\times\Sphere_2$ to the line segments connecting the 
two flats.

Now consider subdividing $\Sphere_1\times\Sphere_2$ 
according to whether the corresponding line segments cross or do not 
cross a hyperplane $\Hyperplane$ of the arrangement.  The 
boundary between crossing and non-crossing line segments is 
formed by the segments with an endpoint on a great
sphere formed by intersecting $\Hyperplane$ with
$\Sphere_1$ or $\Sphere_2$. The line segments that cross $\Hyperplane$
therefore correspond to a set
$(\Halfspace_1\times\overline\Halfspace_2)
\cup(\overline\Halfspace_1\times\Halfspace_2)$,
where $\Halfspace_i$ is a hemisphere bounded by the intersection of
$\Hyperplane$ with $\Sphere_i$.  A line segment crossing the fewest 
hyperplanes then corresponds to a point in the fewest such sets.

For example, Figure~\ref{cover} illustrates the case in which 
$\Flat_1$ and $\Flat_2$
are each lines.
The space of line segments with one endpoint on 
$\Flat_1$ and the other endpoint on $\Flat_2$
is doubly covered by the two-dimensional torus 
$\Sphere_1\times\Sphere_2$, which we have cut along two circles
to show as a square.
The solid dots represent the same line segment;
the hollow dots represent the complementary line segment.
Three covering sets of the form
$(\Halfspace_1\times\overline\Halfspace_2)
\cup(\overline\Halfspace_1\times\Halfspace_2)$
are shown; their boundaries are shown dotted, dashed, and solid
respectively, and the interiors of the sets
are shaded.  (The dotted boundary happens to align
with the circles that cut the torus down to a square.) 

\OCfigure{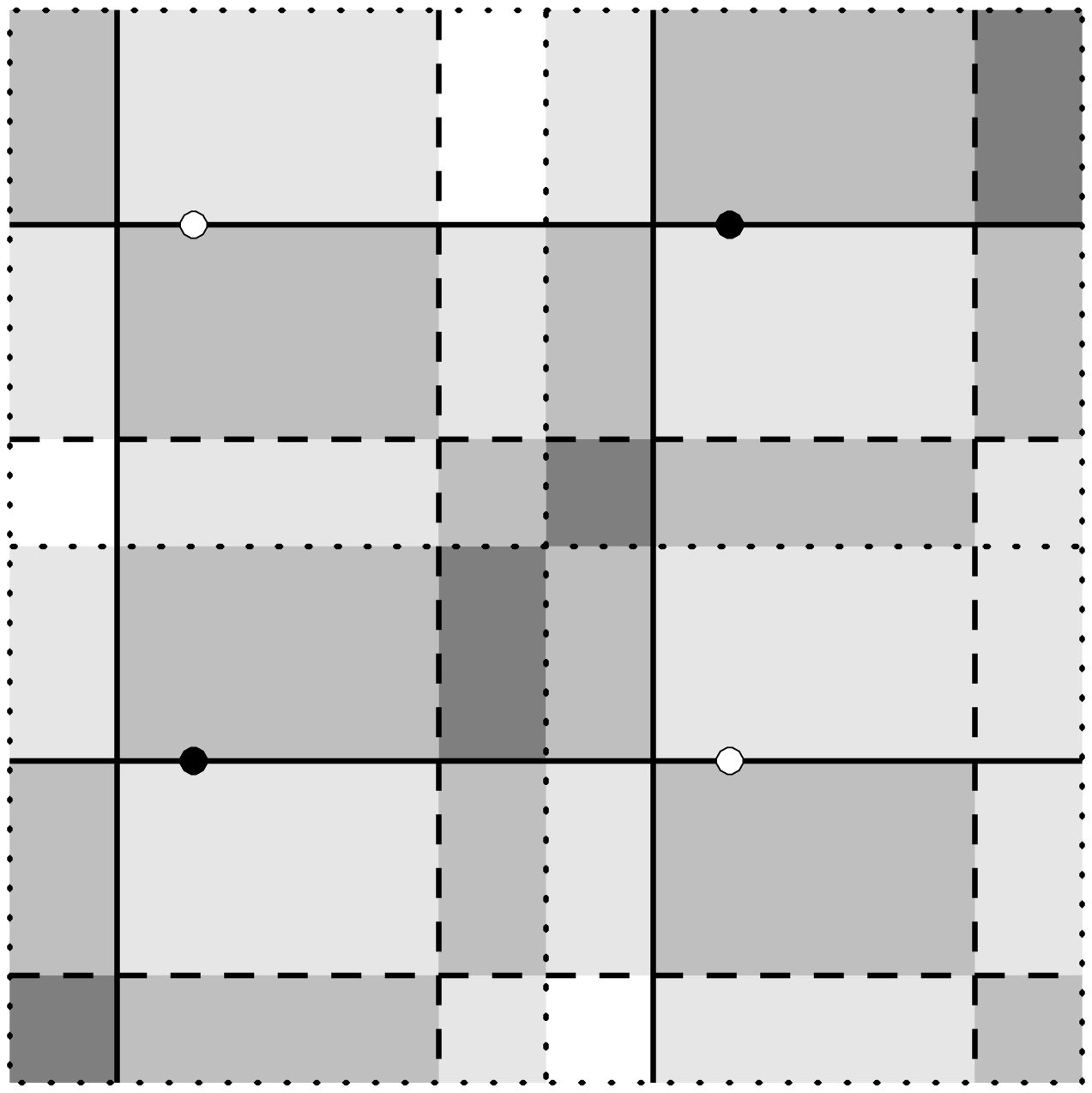}{Computing the crossing distance between
two lines (1-flats) is equivalent to finding a minimally 
covered point on the torus $\Sphere_1 \times \Sphere_2$.}{3.0in}

Since the union in each set of the form
$(\Halfspace_1\times\overline\Halfspace_2)
\cup(\overline\Halfspace_1\times\Halfspace_2)$ is a disjoint union,
we can simplify the problem a bit
by cutting each such set into two products of hemispheres.
We summarize the discussion above with a lemma.

\begin{lemma}\label{deeprep}
Computing the crossing distance between flats 
$\Flat_1$ and $\Flat_2$ is equivalent to finding a point in
a product of spheres 
$\Sphere_1\times\Sphere_2$ that is
covered by the fewest sets from a given family of subsets, each of which
is a product of hemispheres 
$\Halfspace_1\times\Halfspace_2$.
\end{lemma}

\section{Algorithms}

We now show how to solve the problem given in Lemma~\ref{deeprep}.
We first consider the special case of the crossing distance
between a point and hyperplane, that is, $j=0$ and $k=d-1$.
In this case, the product of spheres is a disjoint pair of
$(d-1)$-spheres, both covered identically by a family of hemispheres,
so we can treat it as if it were just a single sphere.
We would like to find a point on this sphere that is covered by the
fewest hemispheres. We can build the entire arrangement of hemispheres in
time
$O(n^{d-1} + n\log n)$ using a slight modification of an algorithm
for computing a hyperplane arrangement~\cite{Ede-87},
and compute the number of hemispheres covering each cell by stepping
from cell to cell in constant time per step. 
Any minimally covered cell gives a solution.

Next let us consider the special case of the crossing distance
between two lines, that is, $j=k=1$.  The product of spheres  
$\Sphere_1\times\Sphere_2$ is just a 2-torus, and the
products of hemispheres are just products of semicircles.
We cut the torus into a square as in Figure~\ref{cover};
each product of semicircles turns into a set of at most four rectangles.
We refer to the horizontal and vertical projections of these 
rectangles as {\it segments\/}.

We can now use a standard sweep-line algorithm to compute a point in 
$\Sphere_1\times\Sphere_2$ covered by a minimum number of sets.
Conceptually we sweep a vertical line from left to right 
across Figure~\ref{cover}.
We use a segment tree~\cite{PreSha-85}
to represent the vertical segments crossed by
the sweep line; let us assume that vertical represents $\Sphere_2$.
As usual with segment trees,
each vertical segment $\Halfspace_2$ 
appears at $O(\log n)$ nodes of the segment tree, exactly those 
nodes whose intervals are covered by $\Halfspace_2$
but whose parents' intervals are not covered by $\Halfspace_2$.
(Here we denote vertical segments by $\Halfspace_2$,
even though some of them are just ``halves'' of the
original semicircles $\Halfspace_2$.)

We equip each node $v$ of the segment tree with an additional
piece of information:  the minimum number of $\Halfspace_2$ segments
covering some point in the interval corrresponding to $v$.
This {\it coverage number\/} can be computed by taking the
minimum of the numbers at $v$'s two children and adding the number of
segments listed at $v$ itself. 

We sweep horizontally across the square.
The {\it events\/} in the sweep algorithm  
correspond to endpoints of segments on $\Sphere_1$.
At each endpoint of a segment we update the segment tree
along with the coverage numbers at its nodes.  
Coverage numbers change at only $O(\log n)$ nodes: the
ancestors of the nodes storing the newly inserted 
or deleted vertical segment. 
The coverage number at the root gives the minimally
covered cell currently crossed by the sweep line.
We also maintain the overall minimum covering seen so far,
and update this minimum at each event.
At the end of the sweep, the overall minimum gives the answer.
We have obtained the following theorem.

\begin{theorem}
The crossing distance between two lines in an arrangement
in $\R^d$, or the regression depth of a line in $\R^3$, can be found
in time $O(n\log n)$.
\end{theorem}

For general $j$ and $k$, we use a randomized recursive
decomposition in place of the segment tree.

\begin{lemma}\label{bsp}
Given an arrangement of hyperplanes in $\R^d$, we can 
produce a recursive binary decomposition of $\R^d$, with
high probability in time $O(n^{d}+n\log{n})$,
such that
any halfspace bounded by an arrangement hyperplane has
(with high probability) a representation as a disjoint union of
decomposition cells with $O(n^{d-1}+\log{n})$ ancestors.
\end{lemma}

\begin{proof}
We apply a randomized incremental arrangement 
construction algorithm. Each cell in the recursive decomposition is
an  arrangement cell at some stage of the construction. The bound on the 
representation of a halfspace comes from applying the methods 
of \cite[pp. 120--123]{Mul-94} to the zone of the boundary hyperplane.
\end{proof}

The same method applies essentially without change to
spheres and hemispheres, so we can apply it to the sets occurring in 
Lemma~\ref{deeprep}.
Each product of hemispheres occurring in Lemma~\ref{deeprep}
can be represented as 
disjoint unions of $O(n^{\,j+k-2})$ products of cells in the product of 
the two recursive decompositions formed by applying 
Lemma~\ref{bsp} to $\Sphere_{1}$ and  $\Sphere_{2}$.
Since there are $O(n)$ products of hemispheres, we have overall
$O(n^{\,j+k-1})$ products of cells.

The algorithm has a similar structure to the algorithm
for the case $j=k=1$, only the simple sweep order for
processing the cells of $\Sphere_1$ is replaced by a
depth-first traversal of the recursive decomposition of $\Sphere_1$.
As in the sweep algorithm, we maintain a 
coverage number for each cell of the
decomposition of $\Sphere_2$. 
The coverage number measures the fewest
$\Halfspace_2$ hemispheres covering some point in that cell, 
where the $\Halfspace_2$ hemispheres
come from pairs $\Halfspace_1\times\Halfspace_2$
for which $\Halfspace_1$ covers the current cell in
the traversal of $\Sphere_1$.
These numbers are computed by taking the
minimum number for the cell's two children and adding the number of
hemispheres whose decomposition uses that cell directly.
When the traversal visits a cell in $\Sphere_{1}$,
we determine the set of hemispheres whose decomposition uses that cell,
and update the numbers for the ancestors of cells covering the
corresponding hemispheres in $\Sphere_2$.
Each hemisphere product leads to $O(n^{\,j+k-2})$
update steps, so the total time for this 
traversal is $O(n^{\,j+k-1})$.  We also maintain the overall minimum
covering seen so far, and take the minimum with the
number at the root of the decomposition of $\Sphere_2$ whenever
the depth-first traversal reaches a leaf in the decomposition of $\Sphere_1$.

When one flat---say $\Flat_{1}$---is a 
line, this method's time includes an unwanted
logarithmic factor.  To avoid this factor, we return 
to a sweep algorithm as the case $j=k=1$.
We sweep across $\Sphere_{1}$,
using the hierarchical decomposition data structure for 
$\Sphere_{2}$ in place of the segment tree.
When the traversal reaches an endpoint of an interval
$\Halfspace_1$, we update the cells for the corresponding hemisphere
$\Halfspace_2$.

We summarize with the following theorem. 
It is likely that $\epsilon$-cuttings can
derandomize this result.

\begin{theorem}
The crossing distance between a $j$-flat and a $k$-flat can be found with
high  probability in time $O(n^{\,j+k-1}+n\log n)$
for $1 \leq j,k$.  The depth of a
$k$-flat for $1 \leq k \leq d-2$ can be found in time
$O(n^{d-2}+n\log n)$ with high probability.
\end{theorem}

\let\section\subsection
\bibliographystyle{abuser}
\let\oldbib\thebibliography
\def\thebibliography#1{\oldbib{#1}\small\itemsep 0pt}
\bibliography{regdepth}
\end{document}